\documentclass[aps,nofootinbib,amsmath,floatfix]{revtex4}
\usepackage{CJK}
\usepackage{amsmath}
\usepackage{amsfonts}
\usepackage{amssymb}
\usepackage{color}
\usepackage{undertilde}

\newcommand{\omits}[1]{}

\definecolor{dyellow}{rgb}{1.,0.8,.0}
\definecolor{myblue}{rgb}{.1,.1,.7}
\definecolor{dcyan}{rgb}{.0,.6,.6}
\definecolor{dmagenta}{rgb}{0.6,0.0,0.6}
\definecolor{brown}{rgb}{0.6,0.2,0.}
\definecolor{darkblue}{rgb}{.0,.0,0.5}
\definecolor{darkred}{rgb}{0.75,0.0,0.0}
\definecolor{orange}{rgb}{1.,.6,.0}
\definecolor{dorange}{rgb}{0.8,.4,.0}
\definecolor{darkgreen}{rgb}{0.0,0.6,0.0}
\definecolor{purple}{rgb}{.4,.0,.4}
\definecolor{grey}{rgb}{0.7,0.7,0.7}

\newcommand{\vecto}[1]{\mbox{\boldmath $#1$}}

\begin{document}
\begin{CJK*}{GBK}{song}

\title{New Geometry with All Killing Vectors Spanning the Poincar\'e Algebra}

\author{HUANG Chao-Guang $^{1,2}$\footnote{Email: huangcg@ihep.ac.cn},
{TIAN Yu $^{3}$\footnote{Email: ytian@gucas.ac.cn}},
{WU Xiao-Ning $^{4,5}$\footnote{Email: wuxn@amss.ac.cn}},
{XU Zhan $^6$\footnote{Email: zx-dmp@tsinghua.edu.cn}},
{ZHOU Bin $^7$\footnote{Email: zhoub@bnu.edu.cn}}}

\affiliation{${}^1$Institute of High Energy Physics, Chinese Academy of
Sciences, Beijing 100049 \\
${}^2$Theoretical Physics Center for Science Facilities, Chinese Academy of
Sciences, Beijing 100049 \\
${}^3$Graduate University of Chinese Academy of
Sciences, Beijing 100049 \\
${}^4$ Institute of Mathematics, Academy of Mathematics and System Sciences,
Chinese Academy of Sciences, Beijing 100190\\
${}^5$ Hua Loo-Keng Key Laboratory of Mathematics, Chinese Academy of Sciences, Beijing
100190\\
${}^6$ Department of Physics, Tsinghua University, Beijing
100084\\
${}^7$ Department of Physics, Beijing Normal University, Beijing
100875}

\begin{abstract}
\noindent {\it The new 4D geometry whose Killing vectors span the Poincar\'e algebra
is presented and its structure is analyzed.  The new geometry can be regarded
as the Poincar\'e-invariant solution of the degenerate extension of
the vacuum Einstein field equations
with a negative cosmological constant and provides
a static cosmological space-time with a Lobachevsky space.  The motion
of free particles in the space-time is discussed.}

\medskip


\end{abstract}


\maketitle


The Poincar\'e symmetry is the foundation of Einstein's
special relativity, relativistic field
theories in Minkowski space-time, particle physics, as well as the Poincar\'e gauge
theories of gravity, etc.  It is well known
that conventionally only the Minkowski space-time is invariant under
global Poincar\'e transformations.

Recently, it has been found that, similar to kinematic symmetry
\cite{BLL}, there is another Poincar\'e symmetry that preserves the
Minkowski lightcone at origin \cite{GWZ,GHWZ}.

In this Letter, we present a new nontrivial four-dimensional (4D) geometry with
the new kinematic Poincar\'e symmetry and to explore its geometric structure.  We show
that the new geometry is the solution of the
degenerate extension of the vacuum Einstein
field equations with a negative cosmological constant and provides a
static cosmological space-time with a Lobachevsky space. The motion
of free particles in the space-time is also discussed.

In the algebraic point of
view, an algebra is said to be Poincar\'e one if \\
(1) it is isomorphic to $\frak{iso}(1,3)$ algebra; \\
(2) \parbox[t]{15.2cm}{the unique Abelean ideal of the $\frak{iso}(1,3)$ algebra
is regarded as a translation sub-algebra and is divided into the time
translation and space translations as a 1D and a 3D representation,
respectively, of $\mathfrak{so}(3)$ sub-algebra of the
$\frak{so}(1,3)$ sub-algebra; and} \\
(3) \parbox[t]{15.2cm}{the algebra is invariant under the
suitable parity and time-reversal operations defined in Ref. \cite{BLL}.}\bigskip

Clearly, in the above algebraic sense, the generator set
$(H, P_i, K_i, J_i)$ with %
\begin{eqnarray}\label{p}
\begin{array}{l}
H = \partial_t ,\ P_i = \partial_i, \smallskip \\
K_i = (t \partial_i +\dfrac 1 {c^2} x^i \partial_t), \ %
J_i = \epsilon_{ij}^{\phantom{ij} k} x^j\partial_k . 
\end{array}
\end{eqnarray}
spans a Poincar\'e algebra ${\frak p}$
of the ordinary Poincar\'e transformation group%
\begin{eqnarray}
  {x'}^\mu = L^\mu_{\ \nu}x^\nu+ l a^\mu, \qquad L\in SO(1,3),
\end{eqnarray}%
where $l$ is a constant having dimension of length, and $a^\mu$
are dimensionless parameters.

Remarkably, there exists another generator set
$(H',P '_i, K_i,J_i)$,
\begin{eqnarray}\label{np2}
\begin{array}{l}
H'  = - c^2 l^{-2} t  x^\kappa \partial_\kappa  \ (=c P_0'), \quad
 P'_i  = l^{-2} x^i  x^\kappa \partial_\kappa ,  \smallskip\\
K_i =   (t \partial_i +\dfrac 1 {c^2} x^i \partial_t), \quad
 J_i = \epsilon_{ij}^{\ \ k} x^j\partial_k ,
\end{array}
\end{eqnarray} %
which also spans a Poincar\'e algebra in the above sense \cite{GWZ, GHWZ},
\begin{equation}
\begin{array}{l}
\ [ H', P'_i ]=0, \qquad \quad \,
[H', K_i]= P'_i, \qquad \quad [H', J_i] = 0,\smallskip \\
\ [ P'_i, P'_j] = 0, \qquad\quad \, [ P'_i, K_j]= \dfrac 1 {c^2} H'\delta_{ij},
\quad [P'_i, J_j]= -\epsilon_{ij}^{\ \ k}P'_k, \smallskip \\
\
[ K_i, K_j ] =  \dfrac 1 {c^2}\epsilon_{ij}^{\ \ k} J_k,
\, [K_i, J_j] =-\epsilon_{ij}^{\ \ k}K_k, \ \quad [J_i, J_j]= -\epsilon_{ij}^{\ \ k} J_k ,
\end{array}
\end{equation}
where the indexes are lowered and raised by $\eta_{\mu\nu}={\rm diag}(1,
-1,-1,-1)$ or $-\delta_{ij}$.
They are the generators of the Poincar\'e transformation group \cite{UWFH}
\begin{eqnarray}\label{CoordTrans}
{x'}^\mu = \dfrac {L^\mu_{\ \nu}x^\nu}{1+l^{-1}b_\mu x^\mu}
\end{eqnarray}
with dimensionless parameters $b_\mu$.

It should be noted that the new generator set does not generate the isometry of
the Minkowski space-time.  Instead, it preserves all straight lines (including
timelike, null, and spacelike ones) through the origin in the Minkowski
space-time.  Therefore, the Poincar\'e group  with the new generator
set is referred to as the second realization of Poincar\'e group
(or the second Poincar\'e group, for brevity,) and denoted as $P_2$
in the following.\footnote{The group defined here
should be
distinguished from the group defined by Aldrovandi {\it et al}
\cite{AP,ABCP}.}
The transformations generated by
the new set of generators are called the second Poincar\'e
transformations, $H'$ and $P'_i$ are named
the pseudo-translations \cite{GWZ, GHWZ} or ${\frak p}_2$-translations.

It is natural to ask: is there a 4D
geometry which is invariant under the second Poincar\'e
transformations?

The following no-go theorem answers the question in conventional way.
{\it There is no tensor field $\vecto{g} = g_{\mu\nu} \, dx^\mu \otimes dx^\nu$
with the following three conditions satisfied simultaneously:\\
(1) $\vecto{g}$ is smooth; \\
(2) $\vecto{g}$ is non-degenerate everywhere; and\\
(3) $\vecto{g}$ is invariant under the $\mathfrak{p}_2$-translations.}
The proof of the theorem can be found in Ref. \cite{HTWXZ}.

It implies that the  metric invariant under
the $P_2$ transformations ($P_2$-invariant, for brevity,) on the 4D underlying
manifold must be
degenerate if it exists. For a degenerate geometry, more geometric information
should be assigned.

It can be checked that $(M,\vecto{g},\vecto{h},\nabla)$ is invariant under $P_2$
transformations, where $\vecto{g}$ is a 4D type-(0,2) degenerate symmetric
tensor field
\begin{eqnarray} \label{g1}%
\vecto{g} &=& g_{\mu\nu}dx^\mu \otimes dx^\nu \nonumber \\
&=&  \frac {l^2} {(x \cdot x)^2}(\eta_{\mu\nu}\eta_{\rho \tau}
-\eta_{\mu \rho}\eta_{\nu \tau})x^\rho x^\tau dx^\mu dx^\nu,
\end{eqnarray}
$\vecto{h}$ is a  4D type-(2,0) degenerate symmetric tensor field
\begin{eqnarray}
\label{gp2inv} \vecto{h}=h^{\mu\nu}\partial_\mu \otimes \partial_\nu =
l^{-4}(x\cdot x)x^\mu x^\nu\partial_\mu \partial_\nu
\end{eqnarray}%
with $x \cdot x=\eta_{\mu\nu} x^\mu x^\nu >0$,
and $\nabla$ is a connection compatible to ${\vecto g}$ and ${\vecto h}$, i.e.
\begin{eqnarray}
\nabla_\lambda \, g_{\mu\nu}= \partial_\lambda g_{\mu\nu}-\Gamma^\kappa_{\lambda\nu}g_{\mu\kappa}
-\Gamma^\kappa_{\mu\lambda}g_{\kappa\nu} =0
\end{eqnarray}
and
\begin{eqnarray}
\nabla_\lambda \,
{h}^{\mu\nu}= \partial_\lambda {h}^{\mu\nu}+ \Gamma^\nu_{\lambda\kappa}{h}^{\mu\kappa}
+\Gamma^\mu_{\lambda\kappa}{h}^{\kappa\nu}=0,
\end{eqnarray}
respectively, with connection coefficients,
\begin{eqnarray}
\label{inv-connectn}%
 \Gamma^\mu_{\ \nu\lambda} = - \dfrac {(x_\nu
\delta^\mu_\lambda+\delta^\mu_\nu x_\lambda) }{x\cdot x}.
\end{eqnarray}
In other words, $\forall \vecto\xi \in {\frak p}_2\subset \Gamma(TM)$,
\begin{eqnarray} \label{eq:Lieg}%
&{\cal L}_{\vecto \xi} \vecto{g} = g_{\mu\nu,\lambda}\xi^\lambda+g_{\mu\lambda}\partial_\nu\xi^\lambda
+g_{\lambda\nu}\partial_\mu\xi^\lambda=0,&\\
\label{eq:Lieh} %
&{\cal L}_{\vecto \xi} {\vecto h}={h}^{\mu\nu}_{\ \ ,\lambda}\xi^\lambda -
h^{\mu\lambda}\partial_\lambda\xi^\nu - h^{\lambda\nu}\partial_\lambda\xi^\mu=0,&\\
\label{eq:LieGamma}%
& [{\cal L}_{\vecto \xi}, \nabla ] =0,&
\end{eqnarray}
are valid simultaneously, or Eqs.(\ref{g1}), (\ref{gp2inv}), and
(\ref{inv-connectn}) are invariant under the coordinate transformation
(\ref{CoordTrans}) and its inverse transformation,
\begin{eqnarray} \label{InverseTrans}
x=\dfrac {L^{-1} x'}{1-l^{-1}(b\cdot L^{-1}x')}=\dfrac {L^{-1} x'}{1-l^{-1}(b'\cdot x')}.
\end{eqnarray}
The curvature tensor of the space-time is, by definition,
\begin{eqnarray}
R^\sigma_{\ \mu \nu \rho}&=&\partial_\nu\Gamma^\sigma_{\ \mu \rho}-
\partial_\rho \Gamma^\sigma_{\ \mu  \nu} +
\Gamma^{\sigma}_{\ \tau \nu} \Gamma^\tau_{\ \mu \rho}-\Gamma^{\sigma}_{\ \tau \rho}
\Gamma^\tau_{\ \mu \nu}\nonumber \\
&=&l^{-2}(\delta^\sigma_\nu g_{\mu\rho}-\delta^\sigma_\rho g_{\mu\nu}).
\label{curvaturecomponents}
\end{eqnarray}
It is antisymmetric in the latter two indexes and satisfies the Ricci and
Bianchi identities. The Ricci curvature tensor is then
\begin{eqnarray} \label{riccicurv}
R_{\mu\nu}=R^\sigma_{\ \mu\nu\sigma}=  -3l^{-2}g_{\mu\nu}.
\end{eqnarray}
They are obviously invariant under $P_2$ transformations.
Eqs.(\ref{curvaturecomponents}) and
(\ref{riccicurv}) are similar to those of the non-degenerate maximum-symmetric
space-times.

In order to see the manifold more transparently,
let us consider the coordinate transformations,
\begin{eqnarray} \label{Transf2AdS}
\begin{array}{l}
x^0=l^2\eta^{-1} \cosh (r/l) ,  \\
x^1=l^2\eta^{-1} \sinh (r/l) \sin \theta \cos\phi ,\\
x^2=l^2\eta^{-1} \sinh (r/l) \sin \theta \sin \phi ,\\
x^3=l^2\eta^{-1} \sinh (r/l) \cos \theta .
\end{array}
\end{eqnarray}
Under the coordinate transformation, Eqs.(\ref{g1}), (\ref{gp2inv}), and
(\ref{inv-connectn}) become
\begin{eqnarray} \label{g3}
{\vecto g} =-(dr^2 +l^{2}\sinh^2(r/l) d\Omega_2^2),
\end{eqnarray}
\begin{eqnarray}\label{h1}
{\vecto h}= (\partial_\eta)^2,
\end{eqnarray}
\begin{eqnarray} \label{connect}\begin{array}{l}
{\bar \Gamma}^\eta_{ij} = -l^{-2}\eta g_{ij} \\
{\bar \Gamma}^r_{\theta\theta} =-l\sinh(r/l) \cosh(r/l), \quad
{\bar \Gamma}^r_{\phi\phi} ={\bar \Gamma}^r_{\theta\theta}\sin^2\theta   \\
{\bar \Gamma}^\theta_{r\theta} ={\bar \Gamma}^\theta_{\theta r} ={\bar \Gamma}^\phi_{r\phi}
={\bar \Gamma}^\phi_{\phi r}
=\dfrac 1 {l\tanh(r/l)} \\
{\bar \Gamma}^\theta_{\phi\phi} =-\sin\theta\cos\theta, \quad
{\bar \Gamma}^\phi_{\theta\phi}={\bar \Gamma}^\phi_{\phi\theta}=\cot\theta  \\
\mbox{others vanish,}
\end{array} %
\end{eqnarray}
where an over bar represents that the quantity takes the value
in the coordinate system $\{{\bar x}^0, {\bar x}^i\}=\{\eta, r,\theta, \phi\}$.
Clearly, Eqs. (\ref{g3}) and (\ref{h1}) define the non-degenerate
3D metric ${\vecto g}_3$ and 1D contravariant metric ${\vecto h}_1$, respectively.
The above structures show that the manifold is locally
$\mathbb{R} \times \mathbb{H}_3 $, where 3D hyperboloid $\mathbb{H}_3 $
has unit `radius'.  Under the $P_2$-transformations (\ref{CoordTrans}),
points run on $\mathbb{R} \times \mathbb{H}_3 $.  The points satisfying
$1+l^{-1} b\,\cdot x =0$
will be transformed to infinity in the new coordinate system $x'$ under the
coordinate transformation (\ref{CoordTrans}).  In particular, when
$L^\mu_{\ \nu}=\delta^\mu_\nu$, they correspond to the points
being transformed from $(\eta,r,\theta,\phi)$ to $(0, r, \theta, \phi)$.
Eqs(\ref{g3}), (\ref{h1}), (\ref{connect}) and the above transformation property show the
geometric structure
can be smoothly extended to $\eta=0$ and $\eta<0$.
Therefore, the manifold is $\mathbb{R} \times \mathbb{H}_3 $ globally.

The Ricci curvature (\ref{riccicurv}) reads in the new coordinate system
\begin{eqnarray}
\bar R_{\mu\nu} = 3l^{-2} {\rm diag}(0, 1, \sinh^2(r/l), \sinh^2(r/l) \sin^2\theta).
\end{eqnarray}
Compared
with the reduced vacuum Einstein field equation with a cosmological constant
\begin{eqnarray} \label{Eeq}
R_{\mu\nu}=\Lambda g_{\mu\nu},
\end{eqnarray}
the new geometry may be regarded as the $P_2$-invariant solution of the
degenerate extension of the vacuum Einstein field equation with the
cosmological constant $-3l^{-2}$.
The solution provides a static cosmological space-time with a Lobachevsky space.

Recall that the Ashtekar extension of general relativity admits the degenerate
geometries \cite{Ash,DG}, which satisfy the gauge, vector, and scalar constraints.
For the solution, the non-zero components of the densitized triad and Ashtekar
connection\footnote{In this case, the
extrinsic curvature has no good definition. } $A^i_a=\Gamma^i_a$  read
\begin{eqnarray}\begin{array}{l}
\tilde E^r_1 =l^2 \sinh^2(r/l)\sin\theta,  \\
\tilde E^\theta_2 = l\sinh(r/l)\sin\theta,  \\
\tilde E^\phi_3 =l\sinh(r/l),
\end{array}
\end{eqnarray}
and
\begin{eqnarray}\begin{array}{l}
\Gamma^3_\theta=\cosh(r/l), \\
\Gamma_\phi^1=\cos\theta,  \\
\Gamma_\phi^2=-\cosh(r/l)\sin\theta,
\end{array}
\end{eqnarray}
respectively.  The constraints are
\begin{eqnarray}
&&{\cal D}_a\tilde E^a_i=\partial_a\tilde E^a_i-\epsilon_{ij}^{\ \ k}\Gamma_a^j\tilde E^a_k=0  \\
&&F_{ab}^i\tilde E^a_i=0,  \\
&&\epsilon_{i}^{\ jk}F^i_{ab}\tilde E^a_j \tilde E^b_k=l^{-2}\epsilon^{ijk}\utilde{\eta}_{abc}\tilde E^a_i
\tilde E^b_j \tilde E^c_k,
\end{eqnarray}
where
\begin{eqnarray}
F^i_{ab}=2\partial_{[a} \Gamma_{b]}^i-\epsilon^{i}_{\ jk}\Gamma_a^j \Gamma_b^k.
\end{eqnarray}
Remember that the indexes are lowered and raised by $\eta_{\mu\nu}
={\rm diag}(1,-1,-1,-1)$ and thus $-\delta_{ij}$.  It is obvious that
\begin{eqnarray}
\begin{array}{l}
\dfrac {d\tilde E^a_i}{d\eta}=0, \\
\dfrac {dA_a^i}{d\eta} =0.
\end{array}
\end{eqnarray}
It again corresponds to the static
solution with $\Lambda=-3l^{-2}$ \cite{CC}.

Furthermore, the 4D volume element on the manifold, defined by
\begin{eqnarray}
\vecto{\epsilon}= l^2\sinh^2(r/l) \sin\theta d\eta\wedge dr\wedge d\theta \wedge d\phi,
\end{eqnarray}
is invariant under the $P_2$-transformations, so the manifold is orientable.
Moreover, the \omits{invariant }tensor ${\vecto h}$ defines an
invariant vector field $\partial_{\eta}$ which is regular on
the whole manifold. Compared with the Newton-Cartan case, it gives an
absolute time direction and, therefore, the space-time is obviously time
orientable.

In addition, one may prove that {\it the obtained covariant
degenerate metric $\vecto{g}$, contravariant degenerate metric
$\vecto{h}$ and the connection $\nabla$ are unique, nontrivial,
compatible ones which are invariant under the $\mathfrak{p}_2$-transformations}.
The detailed discussions are presented in Ref. \cite{HTWXZ}.

If the motion for free particles is still determined by the
geodesic equation%
\begin{eqnarray}
\dfrac {d^2 x^\mu}{d\lambda^2}+\Gamma^\mu_{\
\nu\lambda}\dfrac {dx^\nu}{d\lambda}\dfrac {dx^\lambda}{d\lambda}=0,
\end{eqnarray}
as usual, it gives rise to the `uniform rectilinear' motion%
\begin{eqnarray}
\label{gim}
 x^i= a^i x^0 + lb^i
\end{eqnarray}
with dimensionless constants $a^i$ and $b^i$, in which $x^0$ and
$x^i$ are regarded as the `temporal' and `spatial' coordinates,
respectively.  In the $\mathbb{H}_3$ space,
one may introduce the Beltrami coordinates
$z^i=l {x^i}/{x^0}$.
Then, Eq.(\ref{gim}) reads
\begin{eqnarray}
z^i= \dfrac {b^i} {\cosh (r/l)} \eta + la^i.
\end{eqnarray}
When $r\ll l$, it reduces to $z^i= b^i \eta + la^i$.  This is a
uniform rectilinear motion, because $\eta$ and $z^i$ are time and
spatial coordinates, respectively, in the conventional sense.

The `uniform rectilinear' motion (\ref{gim}) can also be obtained from the
Lagrangian
\begin{eqnarray} \label{Lagrangian}
L= \frac {mlc} {x \cdot x}\sqrt{(\eta_{\mu\nu}\eta_{\eta \tau}
-\eta_{\mu \eta}\eta_{\nu \tau}
)x^\eta x^\tau \dot x^\mu \dot x^\nu}.
\end{eqnarray}
The Euler-Lagrangian equation is equivalent to
\begin{eqnarray}
[(x\cdot x)(\dot x\cdot \dot x)-(x\cdot \dot x)^2 ]
\ddot x_\kappa\ +(\dot x\cdot \ddot x)[(x\cdot \dot x) x_\kappa -(x\cdot x)
\dot x_\kappa] \nonumber\\
+(x\cdot \ddot x)[(x\cdot \dot x)\dot x_\kappa
-(\dot x\cdot \dot x)x_\kappa]=0. \qquad \qquad \qquad\nonumber
\end{eqnarray}
The nonzero determinant of its coefficients for $\ddot x$
requires $\ddot x_\kappa= 0.$

It should be noted that even though $c$ is finite and invariant in the
theory, it does not serves as a limit speed.  It only appears in
$x^0$- or $\eta$-expression to ensure their dimension to be $L$.
Similarly, $l$ is not any limit length though it is finite and invariant.


In conclusion, in addition to the ordinary Poincar\'e
group preserving the metric of Minkowski space-time,  there is the second
Poincar\'e symmetry preserving the geometric structure $(M,\vecto{g},
\vecto{h},\nabla)$. A remarkable feature for the new geometry
$(M,\vecto{g}, \vecto{h},\nabla)$ is that it is somewhat like the
Galilei and Carroll space-times in which the metrics of space and
time split, i.e. one may introduce the non-degenerate metrics for
the space and time separately. Therefore, the new kinematics is
non-relativistic-like in this sense.

From the algebraic point of view, ${\frak p}_2$ algebra has the
$SO(1,3)$ isotropy in which $K_i$ serve as the boost and $(H',
P'_i)$ as the pseudo-translation generators.
The geometrical analysis, however, shows that the generators may have very
different meaning from those appear in the algebra.  In fact, now
the space `translations' on $\mathbb{H}_3$ are generated by
${\cal P}_i=\partial_{z^i} + l^{-2}z_iz^j\partial_{z^j}=\frac {c}{l} K_i-\frac l
{x^0} P'_i$ (where $z_i=-z^i$) instead of $P'_i$, while the boost generators
${\cal K}_i=c^{-1}z_i \partial_{z^0}= \frac l c P'_i$ with $z^0 = \frac {l^2}
{x^0}$.  Therefore,
the `translations'
in the space-time with ${\frak p}_2$ symmetry are no longer generated by the
Abelean ideal in
the algebra and $SO(1,3)$ subgroup is no longer the
isotropy of the space-time at each point. The isotropic group of the
manifold is now $ISO(3)$ and the space-time is thus
the homogeneous space
\begin{eqnarray}
M=ISO(1,3)/ISO(3).
\end{eqnarray}
It is a new
geometry satisfying all three assumptions in Ref. \cite{BLL}.

On the new space-time, the motions of free particles can be well
defined.  The mechanics, field theories and even gravity on the
space-time need to be investigated in order to clarify the
application of the new space-time.  \omits{In fact, it has been shown that the
$P_2$-invariant electromagnetic field equations can be set up in
this space-time.  We shall explore it elsewhere.}

It is straightforward to generalize the 4D degenerate geometries to
higher dimensional degenerate geometries.  On the other hand, in the
higher dimensional theories, there may also be the second Poincar\'e group
as its subgroup of symmetry.  Hence, the geometric structure may appear
in a higher dimension.

Finally, the transformations (\ref{CoordTrans}) are in fact the subset of
the linear
fractional transformations with common denominator.
Thus, it is closely related to the principle of
relativity with two invariant constants \cite{GWZ,GHWZ}.  The relation
of new geometry of the second Poincare group
with these issues and the
physical applications still need to be explored further.

\begin{acknowledgments}\vskip -4mm
We are grateful to late Prof. Guo, who gives us many valuable suggestions.
We would also like to thank Prof. Z.-N. Hu, W. T. Ni and H.-X. Yang and
Dr. H.-t. Wu for their helpful discussion.  The work is supported by the
National Natural Science Foundation of China under Grant Nos. 10775140,
10705048, 10731080, 10975141, 11175245, 11075206 and the Fundamental Research Funds for
the Central Universities under Grant No. 105116.
\end{acknowledgments}

\end{CJK*}

\end{document}